\documentclass[a4paper,11pt]{article}
\usepackage{pos}
\usepackage{orcidlink}
\usepackage{braket}
\usepackage{multirow}
\hypersetup{
    colorlinks=true,
    linkcolor=black,
    filecolor=magenta,      
    urlcolor=cyan,
    citecolor=cyan,
    pdftitle={Overleaf Example},
    pdfpagemode=FullScreen,
    }
\usepackage{layouts}

\DeclareUnicodeCharacter{03A6}{\ensuremath{\Phi}}
\DeclareUnicodeCharacter{03C6}{\ensuremath{\phi}}
\DeclareUnicodeCharacter{03B2}{\ensuremath{\beta}}
\DeclareUnicodeCharacter{0301}{}
\DeclareUnicodeCharacter{207B}{\ensuremath{^{-}}}

\title{Probing Nuclear Structure with Kaonic Atoms through E2 Resonance Mixing}

\author*[a]{S. Manti\orcidlink{0000-0003-3770-0863}}
\affiliation[a]{Laboratori Nazionali di Frascati INFN, Frascati, Italy}

\author[a]{L. De Paolis\orcidlink{0000-0002-4203-9902}}

\author[b,a]{L. Abbene\orcidlink{0000-0001-9633-6606}}
\affiliation[b]{Department of Physics and Chemistry (DiFC), Emilio Segrè, University of Palermo, Palermo, Italy}

\author[a,d]{F. Artibani\orcidlink{0009-0000-8905-3165}}
\affiliation[d]{Università degli Studi di Roma Tre, Dipartimento di Fisica, Roma, Italy}

\author[a]{M. Bazzi\orcidlink{0000-0002-1699-7138}}

\author[e,f,a]{G. Borghi\orcidlink{0000-0001-8488-4728}}
\affiliation[e]{Politecnico di Milano, Dipartimento di Elettronica, Informazione e Bioingegneria, Milano, Italy}
\affiliation[f]{INFN Sezione di Milano, Milano, Italy}

\author[g]{D. Bosnar\orcidlink{0000-0003-4784-393X}}
\affiliation[g]{Department of Physics, Faculty of Science, University of Zagreb, Zagreb, Croatia}

\author[h]{M. Bragadireanu\orcidlink{0009-0001-5217-6003}}
\affiliation[h]{IFIN-HH, Institutul National Pentru Fizica si Inginerie Nucleara Horia Hulubei, 30 Reactorului, 077125, Magurele, Romania}

\author[b,a]{A. Buttacavoli\orcidlink{0000-0002-7188-3651}}

\author[e,f]{M. Carminati\orcidlink{0000-0001-9734-3007}}

\author[a]{A. Clozza\orcidlink{0000-0003-2133-1725}}

\author[a,i]{F. Clozza\orcidlink{0009-0002-3298-0624}}
\affiliation[i]{Università degli Studi di Roma Tor Vergata, Dipartimento di Fisica, Roma, Italy}

\author[j,a]{R. Del Grande\orcidlink{0000-0002-7599-2716}}
\affiliation[j]{Faculty of Nuclear Sciences and Physical Engineering, Czech Technical University in Prague, Břehovà 7, 115 19, Prague, Czech Republic}

\author[a,k,l]{K. Dulski\orcidlink{0000-0002-4093-8162}}
\affiliation[k]{Faculty of Physics, Astronomy, and Applied Computer Science, Jagiellonian University, Kraków, Poland}
\affiliation[l]{Center for Theranostics, Jagiellonian University, Krakow, Poland}

\author[e,f]{C. Fiorini\orcidlink{0000-0002-1157-0143}}

\author[g]{I. Friščić\orcidlink{0000-0002-4743-0572}}

\author[a,\dag]{C. Guaraldo\orcidlink{0000-0002-8923-3438}}

\author[a]{M. Iliescu\orcidlink{0009-0003-3859-5679}}

\author[m]{P. Indelicato\orcidlink{0000-0003-4668-8958}}
\affiliation[m]{Laboratoire Kastler Brossel, Sorbonne Université, CNRS, ENS-PSL Research University, Collège de France, Case 74; 4, place Jussieu, F-75005 Paris, France}

\author[n]{M. Iwasaki\orcidlink{0000-0002-3460-9469}}
\affiliation[n]{RIKEN, Tokyo, Japan}

\author[k,l,a]{A. Khreptak\orcidlink{0000-0002-9482-9770}}

\author[c,o]{J. Marton\orcidlink{0009-0003-1912-285X}}
\affiliation[o]{Atominstitut, Technische Universität Wien, Stadionallee 2, 1020 Vienna, Austria}

\author[k,l]{P. Moskal\orcidlink{0000-0001-5644-5963}}

\author[p]{H. Ohnishi\orcidlink{0000-0002-0615-3214}}
\affiliation[p]{Research Center for Accelerator and Radioisotope Science (RARiS), Tohoku University, Sendai, Japan}

\author[a,q]{K. Piscicchia\orcidlink{0000-0001-6879-452X}}
\affiliation[q]{Centro Ricerche Enrico Fermi, Museo Storico della Fisica e Centro Studi e Ricerche "Enrico Fermi", Roma, Italy}

\author[b,a]{F. Principato\orcidlink{0000-0003-2787-0877}}

\author[a]{A. Scordo\orcidlink{0000-0002-7703-7050}}

\author[a]{F. Sgaramella\orcidlink{0000-0002-0011-8864}}

\author[k]{M. Silarski\orcidlink{0000-0003-2206-0963}}

\author[q,a,h]{D. Sirghi\orcidlink{0009-0002-7486-025X}}

\author[a,h]{F. Sirghi\orcidlink{0000-0002-6143-3200}}

\author[k,l,a]{M. Skurzok\orcidlink{0000-0002-4794-5154}}

\author[a]{A. Spallone\orcidlink{0009-0000-2111-8014}}

\author[p,a]{K. Toho\orcidlink{0009-0001-3245-1418}}

\author[c,\dag]{J. Zmeskal\orcidlink{0000-0003-0815-0639}}

\author[a,h]{C. Curceanu\orcidlink{0000-0002-1990-0127}}

\affiliation[\dag]{Deceased}
\onbehalf{for the SIDDHARTA-2 collaboration}

\emailAdd{Simone.Manti@lnf.infn.it}

\abstract{
Kaonic atoms provide a unique laboratory to investigate the interplay between atomic, nuclear, and strong-interaction physics. In heavy nuclei, atomic transitions can couple to low-lying collective nuclear excitations via the electric quadrupole interaction. When the energy difference between two kaonic atomic levels approaches that of a nuclear $2^+$ excitation, a resonant configuration mixing may occur, known as the E2 nuclear resonance effect. In this work, we investigate the conditions for E2 resonance in kaonic molybdenum isotopes. We describe the mixing using state-of-the-art Dirac–Fock calculations combined with updated nuclear structure inputs, including recent electric quadrupole transition strength values and excitation energies. We evaluate the sensitivity of the effect to key parameters, assess its observability in future experiments such as the EXKALIBUR program, and discuss its impact on cascade dynamics. Our results demonstrate the potential of kaonic atoms as a probe of nuclear structure, complementary to conventional nuclear spectroscopy.
}

\FullConference{
}

\begin{document}
\maketitle
\section{Introduction}
Kaonic atoms, formed by the capture of a negatively charged anti-kaon, provide a unique laboratory to study the interplay between atomic, nuclear, and strong-interaction physics \cite{curceanuKaonicAtomsDAFNE2023a}.
After capture, due to the large kaon mass, the atom is formed in a highly excited configuration and relaxes through an atomic cascade \cite{simonsElectromagneticCascadeChemistry1990}. Subsequent transitions among cascade levels probe different distance scales from the nucleus, allowing the kaon to experience different interactions through several mechanisms \cite{Akylas:1978uj}. In the initial part of the cascade, the kaon interacts with atomic electrons via Auger-type mechanisms, leading to depletion of the electronic shell through electron emission \cite{burbidge_mesonic_1953}. After nearly complete electron depletion, radiative emission becomes dominant, with photon energies in the x-ray range. In the final part of the cascade, where the overlap of the kaon wavefunction with the nucleus is non-negligible, nuclear absorption occurs, leading to the decay of the kaonic atom.
The energies and yields of emitted x-rays provide complementary information on the interactions experienced by the kaon. For high- and medium-$n$ transitions, the energies are accurately described by predictions based solely on quantum electrodynamics (QED). This has been recently demonstrated by the SIDDHARTA-2 experiment \cite{sirghiSIDDHARTA2ApparatusKaonic2024a} at the DA$\Phi$NE collider at the National Laboratories of Frascati (INFN-LNF) \cite{milardi_preparation_2018,milardi_et_dafne_2024,milardi_dane_2021} through measurements of kaonic neon \cite{Sgaramella:2024klx}, used to perform precision tests of QED in bound-state systems \cite{manti2026precision}. X-rays involving low-$n$ transitions start to provide information on strong-interaction effects between the kaon and the nucleus, enabling studies in the quantum chromodynamics (QCD) sector. For instance, in light kaonic atoms such as hydrogen \cite{bazziNewMeasurementKaonic2011}, helium \cite{bazziKaonicHelium4Xray2009}, and deuterium \cite{curceanu2020kaonic}, the lowest atomic states are strongly affected by the kaon--nucleon interaction. Precision measurements of the strong-interaction shift and width of the ground state provide direct experimental constraints on low-energy QCD and chiral effective field theories \cite{Obertova:2022des,obertovaAntikaonAbsorptionNuclear2025a}.
Inferring the strong-interaction shift and width of kaonic-atom levels from x-ray spectroscopy is generally limited by the requirement that the level of interest must be directly in the observed transition \cite{battyMEASUREMENTSTRONGINTERACTION1979b}. In practice, quantities such as the level shift $\epsilon_{nl}$ and width $\Gamma_{nl}$ can only be extracted from x-ray transitions involving that level. This requirement introduces experimental challenges, since detectors must be capable of observing low-$n$ transitions, where the strong interaction effects are largest, while simultaneously providing accurate energy resolution to resolve the corresponding spectral features.
In heavy kaonic atoms, an alternative mechanism allows indirect access to the properties of a level even when it is not directly involved in the observed transition \cite{Leon:1975vu}. This arises from the coupling between atomic transitions and low-lying nuclear excitations, particularly the first $2^+$ states, via the electric quadrupole interaction. When the atomic energy difference is close to a nuclear excitation energy, a near degeneracy enables resonant mixing, so that the atom can effectively de-excite by exciting the nucleus.
This phenomenon, commonly referred to as E2 nuclear resonance mixing \cite{Leon:1975vu}, has previously been successfully investigated in exotic atoms like pionic \cite{Bradbury:1975rw} and antiprotonic \cite{Kanert:1986gf,Klos:2004yp} atoms. The quadrupole coupling modifies the effective decay width of the involved atomic level and consequently alters the cascade dynamics, leading to observable attenuation or redistribution of specific x-ray intensities. This attenuation is directly sensitive to the nuclear width of the lower-lying level, thus providing access to its properties even when it is not directly involved in the observed transition.
The effect of the E2 resonance has also been identified in kaonic atoms, notably for the 6h--5g transition in kaonic molybdenum (KMo) isotopes. Measurements were performed in 1975 \cite{Godfrey:1975vt}, where an attenuation of the 6h--5g line was observed when moving from the near-resonant isotope $^{98}$Mo to isotopes farther from resonance, such as $^{92}$Mo. However, the experimental result was inconclusive because the uncertainty of the attenuation was as large as the reported value of 0.16.
Recently, with the measurement of several kaonic atomic transitions in medium and heavy elements by the SIDDHARTA-2 collaboration, the interest in nuclear resonance effects in kaonic atoms has been renewed. These measurements, performed with high-purity germanium (HPGe) \cite{Bosnar:2024ibq} and cadmium zinc telluride (CZT) detectors \cite{Scordo:2023per}, enable precise studies of kaonic x-ray intensities in the hundreds of keV range, making them particularly suitable for high-$Z$ kaonic atoms and opening the possibility to investigate attenuation mechanisms related to the E2 nuclear resonance effect. In this context, the KAMEO proposal (Kaonic Atoms Measuring nuclear resonance Effects Observables) \cite{DePaolis_2023} aims at dedicated measurements of kaonic molybdenum isotopes to study the E2 nuclear resonance effect. This effort is part of the broader EXKALIBUR program (EXtensive Kaonic Atoms research: from LIthium and Beryllium to URanium) \cite{manti_exkalibur_2025}, which focuses on systematic high-precision spectroscopy of kaonic atoms across a wide range of elements.
In this work, we discuss the relevant quantities needed for the evaluation of E2 nuclear resonance mixing in kaonic atoms. In particular, we analyze the atomic and nuclear parameters that determine the strength of the coupling, including the atomic matrix elements, the energy detuning between atomic and nuclear transitions, and the reduced electric quadrupole transition probabilities. Using relativistic atomic calculations for kaonic atoms, we evaluate the quantities entering the mixing amplitude and discuss their magnitude for the isotopes 92 and 98 in KMo and perspectives for future measurements.
The paper is structured as follows. In Sec.~\ref{sec:e2}, we present the theoretical framework describing the coupling between atomic transitions and nuclear quadrupole excitations in the context of the E2 resonance. In Sec.~\ref{sec:results}, we present sensitivity estimates for Mo isotopes, evaluate the mixing amplitudes, and discuss the prospects for future measurements in KAMEO and EXKALIBUR. Finally, in Sec.~\ref{sec:conclusions}, we summarize our results and discuss possible experimental implications.
%
\section{E2 nuclear resonance mixing in kaonic atoms}\label{sec:e2}
\begin{figure}[h]
    \centering
    \includegraphics[width=1\textwidth]{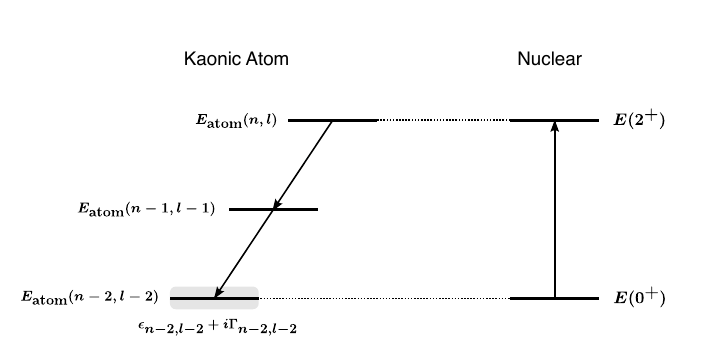}
    \caption{Schematic illustration of the E2 nuclear resonance effect in kaonic atoms. When the energy difference between two kaonic atomic states, $E_{\mathrm{atom}}(n,l) - E_{\mathrm{atom}}(n-2,l-2)$, is close to the energy of a nuclear quadrupole excitation $E(0^+ \rightarrow 2^+)$, a resonant coupling arises via the electric quadrupole interaction. This leads to a mixing between the states  $\ket{n,l;0^+}$ and $\ket{n-2,l-2;2^+}$, with the latter having shift and width $\epsilon_{n-2,l-2} + i\Gamma_{n-2,l-2}$ due to the strong interaction between kaon and nucleus.}
    \label{fig:e2_scheme}
\end{figure}

%
The interaction between the kaonic atom and the nucleus can lead to a coupling between atomic transitions and nuclear excitations when their energies are comparable. In particular, in nuclei with collective quadrupole excitations, the atomic cascade may interact with low-lying $2^+$ nuclear states through the electric quadrupole (E2) interaction \cite{Leon:1975vu}. If the energy difference between two atomic levels is close to the nuclear excitation energy, the atomic transition can resonantly couple to a channel where the nucleus is excited while the kaonic particle transitions to a different atomic state. The process can be understood as the mixing between two configurations: the direct atomic transition $n_i \rightarrow n_f$, which produces the observed x-ray photon, and an alternative channel where the kaonic particle transitions from the initial state $n_i$ to an intermediate atomic state $n_r$ while the nucleus is excited to its first $2^+$ state. The quadrupole interaction imposes a selection rule on the final states, favoring transitions to states with angular momentum $l_r = l_i - 2$. In kaonic atoms, especially for heavy elements, the cascade predominantly proceeds through near-circular states ($l = n-1$), so that the intermediate state is typically also close to circular, with $n_r = n_i - 2$. Therefore, the resonance condition occurs when the energy difference between the atomic levels approximately matches the nuclear excitation energy with:
\begin{equation}
E_\mathrm{atom}(n,l) - E_\mathrm{atom}(n-2,l-2) \simeq E(0^+ \to 2^+).
\end{equation}
Under this condition, the quadrupole interaction between the kaonic particle and the nucleus can induce a mixing between the atomic and nuclear degrees of freedom, modifying the transition amplitude and therefore the probability of the atomic x-ray transition. A schematic representation of the mechanism is shown in Fig.~\ref{fig:e2_scheme}.
The direct atomic transition from the initial level $n-1$ to the final level $n-2$ competes with the resonant channel in which the kaonic particle transitions to the intermediate level $n_r$ while exciting the nucleus from the $0^+$ ground state to the $2^+$ state. The coupling between these two configurations leads to a modification of the effective transition rate of the atomic line. The magnitude of the effect depends on several factors, including the energy detuning between the atomic and nuclear transitions, the reduced electric quadrupole transition probability of the nucleus, and the atomic matrix element associated with the quadrupole interaction. The mixing between the states can be written as:
\begin{equation}
\psi = \sqrt{1-|\alpha|^{2}}\,\ket{n,l;0^{+}} + \alpha\,\ket{n-2,l-2;2^{+}}.
\end{equation}
with the mixing amplitude $\alpha$ defined as:
\begin{equation}
\alpha = \frac{\bra{n-2,l-2;2^{+}} H_{Q} \ket{n,l;0^{+}}}{E_\mathrm{atom}(n-2,l-2)+E(2^{+})-E_\mathrm{atom}(n,l)-E(0^{+})}.
\end{equation}
which can be written in compact form as:
\begin{equation}
\alpha = \frac{\braket{H_Q}}{\Delta - i\Gamma/2}.
\label{eq:alpha}
\end{equation}
where the quadrupole matrix element can be factorized into atomic and nuclear contributions:
\begin{equation}
\braket{H_Q} = \frac{1}{2}\,e^2\,Q_0\,\braket{r^{-3}}\,\mathcal{A}_l
\label{eq:me}
\end{equation}
Here, $Q_0$ is the intrinsic quadrupole moment, $\braket{r^{-3}}$ is the atomic matrix element between $n,l$ and $n-2,l-2$, and $\mathcal{A}_l$ is an angular coefficient written in terms of 3$j$ symbols \cite{ackerStudyChargeDistribution1966}. The quantity $Q_0$ can be obtained from the reduced electric quadrupole transition probability $B(E2;0^+\rightarrow 2^+)$ \cite{bohr_nuclear_1998} through:
\begin{equation}
B(E2;0^+\rightarrow 2^+) = \frac{5}{16\pi}e^2Q_0
\end{equation}
which is tabulated in several databases. In the denominator of Eq.~(\ref{eq:alpha}), the resonance detuning $\Delta$ and width $\Gamma$ are:
\begin{align}
&\Delta = E(0^+ \to 2^+) - E_\mathrm{atom}(n-2,l-2;n,l) + \epsilon_{n-2,l-2} \\
&\Gamma = \Gamma_{n-2,l-2}
\end{align}
where the shift and width of the upper atomic level $n,l$ are negligible with respect to those of the lower level $n-2,l-2$. The mixing leads to an attenuation of the $n,l \rightarrow n-1,l-1$ transition, because the $n,l$ level acquires an induced width related to the width of the lower level:
\begin{equation}
\Gamma_{n,l}^{\mathrm{ind.}} = |\alpha|^2\Gamma_{n-2,l-2}
\end{equation}
The comparison with the experiment is made possible with the attenuation, defined as the ratio of line intensities between a resonant and a non-resonant isotope:
\begin{equation}
A = \frac{I^*(n,l;n-1,l-1)}{I(n,l;n-1,l-1)} \simeq \bigg(1 + \frac{\Gamma_{nl}^{\mathrm{ind.}}}{\Gamma_{n,n-1}^{rad.}}\bigg)^{-1}
\label{eq:att}
\end{equation}
which can be related to the induced width and to the $\Gamma_{n,n-1}^{rad.}$, the radiative rate of the suppressed transition.
To evaluate the mixing and the attenuation, the transition energies, quadrupole integrals and radiative rate were calculated with the Multiconfiguration Dirac--Fock General Matrix Element (\textsc{mcdfgme}) \cite{desclaux1975,indelicato1990} code (v2025.1). The \textsc{mcdfgme} code evaluates transition energies from the total energy difference between the initial and final states, including QED contributions, recoil terms, finite nuclear size effects, and with the possibility of adding residual electron-screening effects through mixed configurations among electrons and a kaon. For the evaluation of kaonic-atom level shifts and widths, we used values obtained from a kaon multinucleon scattering approach, as described in \cite{obertovaAntikaonAbsorptionNuclear2025a}. Nuclear transition energies and quadrupole matrix elements were taken from the database in \cite{PRITYCHENKO20161}. 
%
\section{Results for KMo and future perspectives}\label{sec:results}
\begin{figure}[h]
    \centering
    \includegraphics[width=0.6\textwidth]{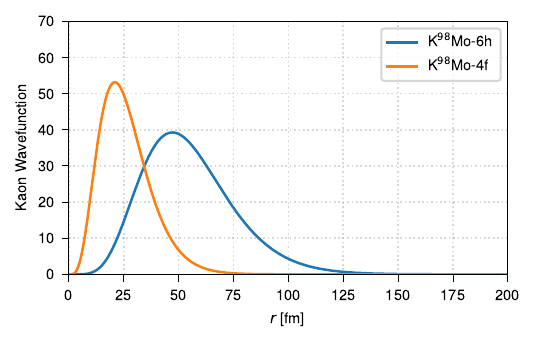}
    \caption{Radial wavefunctions for the kaonic 6h and 4f states in K$^{98}$Mo, shown as the sum of the large and small components as a function of radius in fm. On the scale of this figure, the corresponding wavefunctions for K$^{92}$Mo are indistinguishable, and their differences are therefore not visible.}
    \label{fig:wf}
    
\end{figure}

Here we present results obtained using state-of-the-art atomic calculations and recent nuclear inputs for the kaonic molybdenum isotopes $^{92}$Mo and $^{98}$Mo. We perform calculations with the \textsc{mcdfgme} code for the 6h$\to$5g and 6h$\to$4f transitions in KMo. The latter provides the transition energies and the wavefunctions needed for the integral in Eq.~(\ref{eq:me}), while the former provides radiative rates needed for the attenuation. Fig. \ref{fig:wf} shows the kaon wavefunctions for the states coupled by the E2 resonance, namely 6h and 4f, used for evaluation of the quadrupole matrix element in (\ref{eq:me}). Table \ref{tab:mo_resonance_i} reports all relevant input parameters entering the evaluation of the mixing coefficient $\alpha$, following the formalism introduced in Sec.~\ref{sec:e2}. These include atomic transition energies, nuclear excitation energies, and $Q_0$ values that determine the electric quadrupole coupling matrix element.
%
%
\begin{table}[htbp]
\centering
\caption{Input quantities for the mixing $\alpha$: the transition energy $E_\mathrm{atom}(6h-4f)$, shift $\epsilon_{4f}$ and width $\Gamma_{4f}$ for the 4f level, nuclear excitation energy $E(0^+\to2^+)$ and quadrupole strength $B(E2;0^+\to2^+)$.}
\label{tab:mo_resonance_i}
\begin{tabular}{cccccc}
\hline\hline
\multirow{2}{*}{Isotope} &
$E_\mathrm{atom}(6h-4f)$ &
$\epsilon_{4f}$ &
$\Gamma_{4f}$ &
$E(0^+\to2^+)$ &
$B(E2;0^+\to2^+)$
\\
 &
[keV] &
[eV] &
[keV] &
[keV] &
[$e^2$b$^2$]
\\
\hline
$K^{92}$Mo  & 808.59 & $-12.65$ & 23.88 & $1509.490 \pm 0.030$ & $0.097 \pm 0.006$ \\
$K^{98}$Mo  & 808.84 & $-14.00$ & 27.90 & $787.384 \pm 0.013$ & $0.267 \pm 0.009$ \\
\hline\hline
\end{tabular}
\end{table}

%
The strong-interaction shift and width of the $4f$ level are also reported. The width is of the order of keV, while the corresponding energy shift remains at the level of a few eV, reflecting the different sensitivities of these quantities to the hadron--nucleus interaction.
In Table \ref{tab:mo_resonance_ii}, we report all derived quantities related to mixing and attenuation. As expected, a clear enhancement of the effect is observed for the $^{98}$Mo isotope. This is already evident from the resonance detuning $\Delta$, which shows near-resonant matching between the atomic transition energy and the nuclear excitation energy in this isotope. As a consequence, the denominator in Eq. (\ref{eq:alpha}) is minimized, leading to a larger $\alpha$ and therefore to stronger attenuation of the corresponding x-ray transitions. Conversely, for $^{92}$Mo, the larger detuning suppresses the mixing amplitude, resulting in a smaller $\alpha$ and reduced attenuation. This isotope therefore, provides a useful reference case, where the E2 resonance contribution is present but not dominant. We also report the induced width of the 6h state and the radiative rate of the 6h$\to$5g line, which is nearly independent of the isotope. Combining all results, the corresponding attenuation obtained with Eq.~(\ref{eq:att}) is 0.135$\pm$0.004, in close agreement with the experimental value of 0.16 despite its large uncertainty, and slightly different from previous theoretical estimates of 0.19 because of differences in the input parameters. The uncertainties associated with the nuclear input values are below the percent level, demonstrating excellent sensitivity in this case.
%
\begin{table}[htbp]
\centering
\caption{Values obtained for the resonance detuning $\Delta$, mixing amplitude $|\alpha|$, induced width $\Gamma_{6h}^{ind.}$ and radiative rate $\Gamma_{6h-5g}^{rad.}$.}
\label{tab:mo_resonance_ii}
\begin{tabular}{ccccc}
\hline\hline
\multirow{2}{*}{Isotope} & $\Delta$ & $|\alpha|$ & $\Gamma_{6h}^{ind.}$ & $\Gamma_{6h-5g}^{rad.}$ \\
 & [keV] &  & [eV] & [eV] \\
\hline
$K^{92}$Mo & 700.9 & $0.00092 \pm 0.00003$ & $0.0201 \pm 0.0012$ & 3.30 \\
$K^{98}$Mo & -21.5 & $0.0276 \pm 0.0005$ & $21.20 \pm 0.71$ & 3.31 \\
\hline\hline
\end{tabular}
\end{table}
%

%
The comparison between molybdenum isotopes identifies $^{98}$Mo as a particularly favorable candidate for observing the E2 nuclear resonance effect within the KAMEO program, where dedicated measurements of kaonic Mo isotopes are foreseen. More broadly, the EXKALIBUR program aims to extend such studies to a wide range of heavy elements, including Se, Zr, Ta, Mo, W, and Pb, enabling a systematic investigation of the interplay between atomic and nuclear degrees of freedom in kaonic atoms. Recent experimental developments demonstrate that the required precision is within reach. Within the SIDDHARTA-2 experiment, HPGe detectors have been employed to measure kaonic Pb transitions (9$\rightarrow$8, 8$\rightarrow$7) with sub-5 eV precision, while CZT detectors provide high efficiency, reaching about 70\% at 162 keV, making them well suited for the energy range of heavy kaonic atoms. These capabilities allow precise determination of intensity ratios, which are the key observable for identifying the attenuation induced by E2 mixing, as originally demonstrated in hadronic atoms~\cite{Godfrey:1975vt}. In this context, the development of reliable predictive methods for the mixing amplitude and attenuation is essential. Such tools enable the identification of optimal isotopes and transitions, guiding future measurements and maximizing the sensitivity to E2 resonance effects across different systems. Performing systematic studies across several isotopes is essential to fully exploit the sensitivity of the E2 nuclear resonance effect to nuclear structure. This approach has already been demonstrated in antiprotonic tellurium, where measurements of attenuation allowed access not only to neutron density distributions at the nuclear surface but also to nuclear rms radii \cite{Klos:2004yp}. Extending such studies to molybdenum isotopes is particularly important, given the presence of multiple stable isotopes and the possibility of performing isotope comparisons under controlled conditions. In particular, the isotopes $^{98}$Mo and $^{100}$Mo are directly connected to double-beta decay, making precise information on their neutron distributions and rms radii highly relevant for constraining nuclear matrix elements. These inputs are crucial for the interpretation of neutrinoless double-beta decay, a lepton-number-violating process whose observation would establish the Majorana nature of the neutrino \cite{Agostini:2022zub}.
%
\section{Conclusions and Outlook}\label{sec:conclusions}
In summary, we have presented a comprehensive study of the E2 nuclear resonance effect in kaonic atoms, establishing it as a powerful and complementary tool to probe nuclear structure through kaonic-atom spectroscopy. The E2 resonance mechanism arises from the near-degeneracy between atomic transition energies and low-lying nuclear $2^+$ excitations, leading to configuration mixing between atomic and nuclear channels. As first demonstrated in hadronic systems, even a weak quadrupole interaction can induce sizable effects when the detuning is small, resulting in measurable attenuation of specific x-ray transitions. In this context, it is essential to revisit this effect using modern theoretical tools. The use of state-of-the-art \textsc{mcdfgme} calculations, combined with updated nuclear databases for excitation energies and quadrupole strengths, allows a reliable evaluation of the sensitivity of the effect to both atomic and nuclear inputs, significantly improving upon earlier studies. As a concrete case, we have investigated kaonic molybdenum isotopes, focusing on the attenuation of the $6h \rightarrow 5g$ transition as a reference and the impact of the resonance on lower transitions. The analysis highlights how the proximity between atomic level spacings and nuclear excitation energies enhances the mixing, leading to observable modifications of the cascade intensities. We have evaluated the relevant quantities entering the mixing amplitude and quantified the expected attenuation, showing that the effect is particularly enhanced in $^{98}$Mo due to its favorable detuning. These results demonstrate that measurable deviations in intensity ratios are within reach of current experimental capabilities. These findings are directly relevant for upcoming experimental efforts. Dedicated measurements within the KAMEO program, together with the broader EXKALIBUR initiative, could enable systematic investigations of E2 resonance effects across a wide range of medium and heavy nuclei, providing new opportunities to study the interplay between atomic and nuclear degrees of freedom. Beyond the specific case of molybdenum, the sensitivity of the attenuation to nuclear excitation energies and widths opens a novel pathway to constrain nuclear structure quantities, including those relevant for neutron distributions and the modeling of double-beta decay matrix elements. Overall, kaonic atoms emerge as a unique laboratory to access otherwise inaccessible nuclear properties, offering new insights into hadron--nucleus interactions and establishing a promising direction for future precision studies at the interface of atomic and nuclear physics.
\section{Acknowledgments}
We thank Jaroslava Obertova for providing estimates of the nuclear shifts and widths. We gratefully acknowledge Polish high-performance computing infrastructure PLGrid (HPC Center: ACK Cyfronet AGH) for providing computer facilities and support within computational grant no. PLG/2025/018524. We thank C. Capoccia from INFN-LNF and H. Schneider, L. Stohwasser, and D. Pristauz-Telsnigg from Stefan Meyer-Institut for their fundamental contribution in designing and building the SIDDHARTA-2 setup. We also thank INFN-LNF and the DA$\Phi$NE staff for the excellent working conditions and their ongoing support. Special thanks to Catia Milardi for her continued support and contribution during the data taking. Part of this work was supported by the INFN (KAONNIS project); the Austrian Science Fund (FWF): [P24756-N20 and P33037-N]; the Croatian Science Foundation under the project IP-2022-10-3878; the EU STRONG-2020 project (Grant Agreement No. 824093); the EU Horizon 2020 project under the MSCA (Grant Agreement 754496); the Japan Society for the Promotion of Science JSPS KAKENHI Grant No. JP18H05402, JP22H04917; the Polish Ministry of Science and Higher Education grant No. 7150/E-338/M/2018 and the Polish National Agency for Academic Exchange (grant no PPN/BIT/2021/1/00037); the EU Horizon 2020 research and innovation programme under project OPSVIO (Grant Agreement No. 101038099).
\bibliographystyle{unsrtnat}
\bibliography{KMo_POS}
\end{document}